\documentclass[twocolumn,showpacs,amsmath,amssymb,prd,eqsecnum]{revtex4}

\usepackage{graphicx}
\usepackage{dcolumn}
\usepackage{bm}
\usepackage{epsf}
\usepackage{hyperref}
\usepackage{amsfonts}

\def\Res{\mbox{Res}}

\begin{document}

\title{Black Hole-Neutron Star Binaries in General Relativity:
  Quasiequilibrium Formulation}

\author{Thomas W. Baumgarte}

\altaffiliation{Department of Physics, University of Illinois at
        Urbana-Champaign, Urbana, IL, 61801}

\affiliation{Department of Physics and Astronomy, Bowdoin College,
        Brunswick, ME 04011}

\author{Monica L. Skoge}

\altaffiliation{Present address: Department of Physics, Princeton
University, Princeton, NJ}

\affiliation{Department of Physics and Astronomy, Bowdoin College,
        Brunswick, ME 04011}

\author{Stuart L. Shapiro}

\altaffiliation{Department of Astronomy and NCSA, University of Illinois at
        Urbana-Champaign, Urbana, IL, 61801}

\affiliation{Department of Physics, University of Illinois at
        Urbana-Champaign, Urbana, IL, 61801}

\begin{abstract}
We present a new numerical method for the construction of
quasiequilibrium models of black hole-neutron star binaries.  We solve
the constraint equations of general relativity, decomposed in the
conformal thin-sandwich formalism, together with the Euler equation
for the neutron star matter.  We take the system to be stationary in a
corotating frame and thereby assume the presence of a helical Killing
vector.  We solve these coupled equations in the background metric of
a Kerr-Schild black hole, which accounts for the neutron star's black
hole companion.  In this paper we adopt a polytropic equation of state
for the neutron star matter and assume large black hole--to--neutron
star mass ratios. These simplifications allow us to focus on the
construction of quasiequilibrium neutron star models in the presence
of strong-field, black hole companions.  We summarize the results of
several code tests, compare with Newtonian models, and locate the
onset of tidal disruption in a fully relativistic framework.
\end{abstract}

\maketitle

\section{Introduction}

Black hole-neutron star (hereafter BHNS) binaries have attracted
considerable astrophysical interest recently. For example, they are
candidates for the central engines of short-period, gamma-ray bursts
\cite{jerf99}.  They are also very promising sources of gravitational
radiation for the new generation of ground-based gravitational wave
detectors, including the Laser Interferometer Gravitational Wave
Observatory (LIGO), and the future space-based gravitational wave
detector, the Laser Interferometer Space Antenna (LISA).  Observation
of a BHNS binary -- in particular, the tidal disruption of a neutron
star by a black hole companion -- will provide spectacular information
on the gravitational fields in relativistic objects and the nuclear
physics governing neutron star matter (e.g.~\cite{v00}).  However,
BHNS binaries may be among the most challenging objects in
relativistic astrophysics to model numerically.  Any realistic
treatment must deal simultaneously with the difficulties associated
with the spacetime singularity inside the black hole and the
complexities associated with relativistic hydrodynamics.

The inspiral of BHNS binaries is driven by the emission of
gravitational radiation, which extracts both energy and angular
momentum from the binary and also tends to circularize the orbit.  As
long as the binary separation is large, the inspiral is very slow and
radial velocities are very small compared to the orbital velocities
once the orbit has been circularized.  During this phase, the inspiral
can be modeled as a sequence of quasicircular binary orbits.  The
quasiadiabatic inspiral phase ends when dynamical effects become
important.  For BHNS binaries, there are at least two, very different,
possible termination points: the binary may reach an innermost stable
circular orbit (ISCO), at which point the orbit becomes unstable and
the black hole and neutron star plunge rapidly towards each other,
{\it or}, prior to reaching the ISCO, the neutron star might be
tidally distrupted by the gravitational field of the black hole.
Predicting which one of these scenarios actually happens, and the
binary separation at which it occurs, depends on a number of factors
characterizing the system, including the binary mass ratio.

The importance of tidal effects can be estimated, in a Newtonian
framework, by comparing the tidal force on a test mass $m$ on the
surface of the neutron star
\begin{equation}
F_{\rm tid} \sim G \frac{m M_{\rm BH} R_{\rm NS}}{s^3},
\end{equation} 
with the gravitational force exerted by the star
\begin{equation}
F_{\rm grav} \sim G \frac{m M_{\rm NS}}{R_{\rm NS}^2}.
\end{equation}
Here $M_{\rm BH}$ and $M_{\rm NS}$ are the masses of the black hole
and the neutron star, $R_{\rm NS}$, is the neutron star radius and $s$
is the binary separation.  Equating the two forces yields an
approximate critical ``tidal'' separation
\begin{equation} \label{m_ratio_scaling}
\frac{s_{\rm tid}}{R_{\rm NS}} \sim
\left( \frac{M_{\rm BH}}{M_{\rm NS}} \right)^{1/3},
\end{equation}
or, in gravitational units with $G=c=1$,
\begin{equation} \label{isco_scaling}
\frac{s_{\rm tid}}{M_{\rm BH}} \sim
\left( \frac{M_{\rm NS}}{M_{\rm BH}} \right)^{2/3} 
\frac{R_{\rm NS}}{M_{\rm NS}},
\end{equation}
within which tidal forces begin to dominate and may lead to tidal
disruption.

Neutron stars typically have compaction ratios of $R_{\rm NS}/M_{\rm
NS} \sim 5$.  That means that for neutron stars orbiting {\it
supermassive} black holes, $M_{\rm BH} \gg M_{\rm NS}$, so that the
tidal separation is well inside the ISCO, which resides at about $s
\sim 6 M_{\rm BH}$ (the value for a point-mass orbiting a
Schwarzschild black hole).  It is therefore appropriate to neglect any
effects of tidal distortion or internal structure in calculations of
neutron stars inspiral onto supermassive black holes.  Such binaries
may be a promising sources of low-frequency gravitational radiation
for detection by LISA.

For neutron stars orbiting {\it stellar} mass black holes, however,
the internal structure of the neutron star may become important before
it reaches the ISCO.  Such objects may be important high-frequency
sources for LIGO and other ground based gravitational wave detectors.
Any future detection of such inspiraling binaries may therefore carry
information about the internal structure of the neutron star and its
equation of state \cite{v00}.

Considerable effort has gone into the theoretical modeling of binaries
containing either two black holes (e.g.~\cite{c94,b00,ggb02}) or two
neutron stars (e.g.~\cite{wm95,bcsst,use00,ggtmb01,tg02}; see also the
reviews \cite{bs03,c00}).  For the most part, BHNS binaries have been
studied only in Newtonian gravity (but see \cite{m01}).  Several
authors \cite{lrs93,tn96} have modeled BHNS binaries as Newtonian
ellipsoids around point masses, generalizing the classic Roche model
for incompressible stars (see, e.g. ~\cite{ch69}) to compressible
configurations.  These ellipsoidal calculations have also been
generalized to include relativistic effects \cite{f73,s96,wl00}.
Equilibrium models of Newtonian BHNS binaries also have been
constructed by solving the exact fluid equations numerically and again
treating the black hole as a point mass \cite{ue99}.  Dynamical
simulations of Newtonian BHNS binaries, including coalescence, merger,
and the tidal disruption of the neutron star, have been performed, for
example, by \cite[and references therein]{jerf99,l01}. Dynamical
simulations have also been performed for the tidal disruption of
normal stars orbiting massive black holes (see, e.g. \cite{kl04} and
references therein). But, at best, these simulations typically employ
a relativistic treatment of the hydrodynamics with an approximate
treatment of the gravitational field of the star and/or the
gravitational tidal field of the black hole companion. There is
greater urgency in solving this problem more carefully, now that the
possible detections by the Chandra X-ray Observatory of tidal
disruptions of normal stars by supermassive black holes have been
reported \cite{hg04}.  Clearly, accurate and reliable models of
binaries containing black holes, including BHNS binaries, require a
fully relativistic treatment.

This paper is the first in a series on quasiequilibrium models of BHNS
binaries in full general relativity.  Our models serve a dual purpose:
(1) constant rest-mass, quasiequilibrium sequences parametrized by
binary separation approximate the adiabatic inspiral phase and can be
used to locate the ISCO or the onset of tidal instability, and (2)
individual models provide numerically exact initial data for future
relativistic dynamical simulations.  Similar to earlier treatments of
binary neutron stars (e.g.~\cite{wm95,bcsst,use00,ggtmb01,tg02}) and
binary black holes (e.g.~\cite{ggb02}), we adopt the conformal
thin-sandwich decomposition \cite{y99} of the constraint equations of
general relativity.  We solve these equations for the gravitational
field, together with the relativistic Euler equation for the matter,
assuming the presence of a helical Killing vector.  For the background
geometry, we adopt the metric corresponding to a single black hole in
Kerr-Schild coordinates, thereby accounting for the black hole
companion of the neutron star.

In this first paper we provide solutions for the simplest case in
which the mass ratio is large, i.e., $M_{\rm BH} \gg M_{\rm NS}$.
While the estimate above shows that, in this limit, the internal
structure of the neutron star and tidal distortion is small, it allows
us to formulate the BHNS problem in full general relativity and to
construct numerical models of a neutron star in the presence of a
black hole companion.  We also adopt a polytropic equation of state
for the neutron star matter.  In a future paper we will relax these
approximations and will construct binaries with companions of
comparable mass, for which the internal struture, tidal distortion and
tidal disruption are potentially very important.

The paper is structured as follows.  In Section \ref{sec2} we describe
how the BHNS problem can be solved in Newtonian gravity.  A very
similar approach is adopted in the relativistic treatment, so it is
useful to outline this approach first in the much more transparent
Newtonian framework.  In Section \ref{sec3} we develop the
relativistic formalism.  We introduce the assumption of extreme mass
ratios in Section \ref{sec4}. In Section \ref{sec5} we present
numerical results, including code tests and constant rest-mass
sequences. We provide a brief summary in Section \ref{sec6}.  A
detailed description of the numerical strategy that we adopted in
solving the relativistic equations, together with Kerr-Schild
background expressions that appear in these equations, can be found in
Appendix \ref{appA}.  Throughout this paper we continue to 
adopt geometrized units ($G = c = 1$).

%
%

\section{BHNS Binaries in Newtonian Gravity}
\label{sec2}

Before turning to the relativistic problem, it is useful to lay out in
this Section how BHNS binaries can be constructed in the much simpler
Newtonian framework (compare \cite{ue99}).  In the Sections that
follow we will adopt a very similar strategy and algorithm for the
construction of the same binaries in general relativity, and will
refer back to this Section for clarity.

\subsection{The Poisson Equation}
\label{Npoisson}

In Newtonian gravity, the black hole can be described by a point mass
$M_{\rm BH}$ which gives rise to the gravitational potential
\begin{equation} \label{NphiBH}
\phi_{\rm BH} = - \frac{M_{\rm BH}}{r_{\rm BH}},
\end{equation}
where $r_{\rm BH}$ is the distance from the point mass.  The
total gravitational field $\phi$ can be written as a sum
\begin{equation}
\phi = \phi_{\rm BH} + \phi_{\rm NS},
\end{equation}
where $\phi_{\rm NS}$ is the neutron star contribution to the
potential.  Since $\phi_{\rm BH}$ is a homogeneous solution
to the Poisson equation (for $r_{\rm BH} > 0$), we can write
the Poisson equation as
\begin{equation} \label{poisson1}
D^2 \phi = D^2 \phi_{\rm NS} = 4 \pi \rho_0
\end{equation}
where $D^2$ is the Laplace operator.  

\subsection{The Integrated Euler Equation}
\label{Nbernoulli}

\begin{figure}
\includegraphics[scale=0.4]{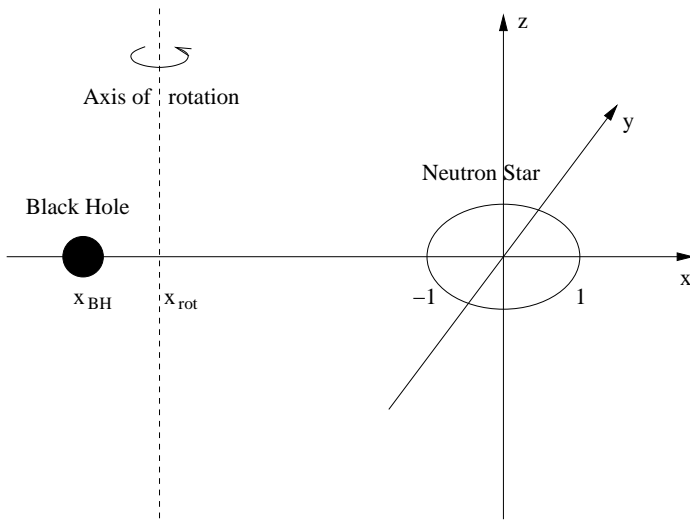} 
\caption{Coordinate system that we use for the construction of BHNS
binaries.  The origin is centered on the neutron star, the black hole
is located at the position $x_{\rm BH}$, and the axis of rotation,
which is parallel to the $z$-axis, is located at $x_{\rm rot}$.  For
the extreme mass ratio $M_{\rm BH} \gg M_{\rm NS}$ adopted in this
paper we have $x_{\rm BH} = x_{\rm rot}$.}
\label{setup}
\end{figure}

We assume a polytropic equation of state
\begin{equation}
P = \kappa \rho_0^{1 + 1/n},
\end{equation}
where $P$ is the pressure, $\rho_0$ the (rest-mass) density, $n$ the
polytropic index and $\kappa$ the polytropic constant.  For corotating
equilibrium configurations the Euler equations can then be integrated
analytically to yield
\begin{equation} \label{bern1}
(n + 1) \frac{P}{\rho_0} + \phi - 
\frac{1}{2} \Omega^2 \left( (x - x_{\rm rot})^2 + y^2 \right) = C,
\end{equation}
where $\Omega$ is the orbital angular velocity, $C$ is a constant and
where we have assumed that the star orbits about an axis parallel to
the $z$-axis located at $x = x_{\rm rot}$ and $y = 0$.  In the
following we will also assume that the black hole and neutron star are
aligned with the $x$-axis (see Fig.~\ref{setup}).

We point out that it is unlikely that the neutron star viscosity could
be strong enough to lock the star into corotation during the binary
inspiral \cite{bc92,visc_fn}.  Therefore it would be more realistic
astrophysically to assume an irrotational rather than a corotational
fluid flow (see, e.g.,~\cite{bs03} and references therein).  However,
the numerical implementation of irrotational fluid flow is much more
involved than that of corotation, and it is therefore reasonable to
focus on corotating configurations first (as it was done for binary
neutron stars).

Equations (\ref{poisson1}) and (\ref{bern1}) are the fundamental
equations describing a stationary star orbiting a black hole in
circular co-rotation.  The solution will depend on three parameters,
which can be chosen, for example, to be the neutron star mass
\begin{equation} \label{mass1}
M_{\rm NS} = \int \rho_0 d^3 x,
\end{equation}
the black hole mass $M_{\rm BH}$ and the separation $d$.  For
computational reasons we will use the maximum density of the neutron
star as one of the three free parameters instead of the neutron star
mass.  Given these parameters, a solution to the problem will also
provide the eigenvalues $\Omega$ and $C$, as well as the size of the
neutron star.

\subsection{Elimination of Dimensions}
\label{Nelimdim}

It is convenient to introduce the dimensionless density parameter
\begin{equation} \label{q}
q = \frac{P}{\rho_0},
\end{equation}
in terms of which we have
\begin{equation}
\rho_0 = \kappa^{-n} q^{n}
\end{equation}
and 
\begin{equation} \label{eos}
P = \kappa^{-n} q^{n+1}.
\end{equation}
Since $\kappa^{n/2}$ has units of length, we can also introduce
dimensionless coordinates $\bar x = \kappa^{-n/2} x$, and similarly
for $y$ and $z$.  The Laplace operator then scales as $\bar D^2 =
\kappa^{n} D^2$, and the angular velocity as $\bar \Omega =
\kappa^{n/2} \Omega$.  In terms of these, equations (\ref{poisson1})
and (\ref{bern1}) become
\begin{equation} \label{poisson2}
\bar D^2 \phi_{\rm NS} = 4 \pi q^n
\end{equation}
and
\begin{equation} \label{bern2}
(n + 1) q + \phi - \frac{1}{2} 
\bar \Omega^2 \left( (\bar x - \bar x_{\rm rot})^2 + \bar y^2 \right) = C.
\end{equation}
The integrated Euler equation (\ref{bern2}) can also be rewritten as
\begin{eqnarray} \label{bern2a}
q & = & \frac{1}{n + 1} \left(C - \phi + \frac{1}{2} 
\bar \Omega^2 \left( (\bar x - \bar x_{\rm rot})^2 + \bar y^2 \right) \right)
	\nonumber \\
& = &  \frac{1}{n + 1} \left(C - \Phi_{\rm eff} \right),
\end{eqnarray}
where the effective potential $\Phi_{\rm eff}$ combines
the gravitational and centrifugal potentials,
\begin{equation} \label{phi_eff}
\Phi_{\rm eff} = \phi -  \frac{1}{2} 
\bar \Omega^2 \left( (\bar x - \bar x_{\rm rot})^2 + \bar y^2 \right).
\end{equation}
This demonstrates that, up to a constant, the density parameter $q$ is
proportional to the effective potential. 

Rescaling $\bar M = \kappa^{-n/2} M$ we also find
\begin{equation}
\bar M_{\rm NS} = \int q^n d^3 \bar x.
\end{equation}

\subsection{Rescaling}
\label{Nrescale}

As we will see below, it is convenient for numerical purposes to have
the surface of the neutron star intersect the $x$-axis at fixed coordinate
locations, say $\hat x_A = -1$ and $\hat x_B = 1$.  This means that we
have to rescale the coordinates with respect to the physical,
non-dimensional size of the neutron star, say $\bar r_e$.  For
spherical stars, $\bar r_e$ is the radius of the star, and otherwise
it is half the star's diameter along the $x$-axis.  We denote these
new coordinates with hatted quantities, $\hat x = \bar x/\bar r_e$, and
similarly for $y$ and $z$.  The derivative operator scales
according to $\hat D_i = \bar r_e \bar D_i$, and we also have $\hat
\Omega = \bar r_e \bar \Omega$.  The Poisson equation (\ref{poisson2})
then becomes
\begin{equation} \label{poisson3}
\hat D^2 \phi_{\rm NS} = 4 \pi \bar r_e^2 q^n
\end{equation}
and the integrated Euler equation (\ref{bern2}) 
\begin{equation} \label{bern3}
(n + 1) q + \phi - \frac{1}{2} 
\hat \Omega^2 \left( (\hat x - \hat x_{\rm rot})^2 + \hat y^2 \right) = C.
\end{equation}
In these coordinates, the physical, dimensionless neutron star mass
$\bar M_{\rm NS}$ can be computed from
\begin{equation}
\bar M_{\rm NS} = \int q^n d^3 \bar x = \bar r_e^3 \int q^n d^3 \hat x. 
\end{equation}

From the Poisson equation (\ref{poisson3}) it is evident that the
gravitational potential $\phi$ will depend on $\bar r_e$.  To make
this dependence explicit, we could introduce a rescaled potential
\begin{equation} \label{scale_phi_NS}
\phi_{\rm NS} = \bar r_e^2 \hat \phi_{\rm NS}.
\end{equation}
This expression can then be inserted into (\ref{bern3}), explicitly
introducing $\bar r_e$ into the integrated Euler equation.  We have
yet to determine the scaling of the black hole contribution $\phi_{\rm
BH}$
\begin{equation} \label{scale_phi_BH1}
\phi_{\rm BH} = \frac{M_{\rm BH}}{r_{\rm BH}}
= \frac{\bar M_{\rm BH}}{\bar r_{\rm BH}}
= \frac{1}{\bar r_e} \, \frac{\bar M_{\rm BH}}{\hat r_{\rm BH}}.
\end{equation}
Here $\bar M_{\rm BH}$ is the dimensionless black hole mass
that we want to keep fixed during the iteration.

With these choices, the integrated Euler equation (\ref{bern3}) can
now be rewritten as
\begin{equation} \label{bern4}
(n + 1) q + \phi_{\rm NS} 
- \frac{\bar M_{\rm BH}}{\bar r_e \hat r_{\rm BH}}
- \frac{1}{2} \hat \Omega^2 \left(
(\hat x - \hat x_{\rm rot})^2 + \hat y^2 \right) = C.
\end{equation}
This equation has to be solved together with the Poisson equation
(\ref{poisson3}).  A self-consistent solution can be constructed
iteratively, as outlined immediately below. 

\subsection{Iteration Scheme}
\label{Niter}

Instead of fixing the neutron star mass, we fix the maximum density on
the $x$-axis $q_{\rm max}$.  A neutron star of a desired mass can
later be constructed iteratively by varying $q_{\rm max}$.  We also
fix the separation $x_{\rm BH}$ in terms of the neutron star size
$\bar r_e$, $\hat x_{\rm BH} = \bar x_{\rm BH} / \bar r_e$.  In the
following we will locate the point mass at $(- \hat x_{\rm BH},0,0)$.
The three input parameters are therefore $q_{\rm max}$, $\hat x_{\rm
BH}$ and $\bar M_{\rm BH}$.

In this paper we also assume that the mass of the black hole is much
greater than that of the neutron star
\begin{equation}
\bar M_{\rm BH} \gg \bar M_{\rm NS},
\end{equation}
in which case the rotation axis will go through the center of the 
black hole
\begin{equation}
\hat x_{\rm rot} = \hat x_{\rm BH}.
\end{equation}
This assumption leads to an error in the location of the rotation axis
of the order of $M_{\rm NS}/M_{\rm BH}$ times the binary separation;
this becomes negligible as the mass ratio decreases.  Without this
assumption, the location of the rotation axis $\hat x_{\rm rot}$ has
to be found as part as the iteration scheme.  Various approaches can
be chosen to identify $\hat x_{\rm rot}$; in Newtonian theory it
coincides with the center of mass.
 
In the limit of extreme mass ratios we expect that the neutron star
will affect the physical solution only in a neighborhood of the star
itself, so that we can confine the numerical grid to a finite domain
around the neutron star that avoids the black hole.  This
significantly simplifies the problem, especially in the relativistic
case, where the black hole interior does not have to be excised from
the numerical grid.

The iteration scheme starts with an initial guess for the density $q$,
confined between $\hat x_A$ and $\hat x_B$, and the star's physical
size $\bar r_e$. 

\subsubsection{Solution of the Poisson equation}

Given a density distribution $q$ we can solve the Poisson equation
(\ref{poisson3}) to find the neutron star potential $\phi_{\rm
NS}$.  The elliptic equation is solved with PETSc algorithms that are
described in more detail in the relativistic context below. 

\subsubsection{Determination of Eigenvalues}

Before the integrated Euler equation (\ref{bern4}) can be solved for
the new matter distribution $q$, the eigenvalues $\hat \Omega$, $\bar
r_e$ and $C$ have to be determined.  This can be done by evaluating
the integrated Euler equation at the three points $\hat x_A$, $\hat
x_B$ and $\hat x_C$.  The latter is defined as the point along the
$x$-axis at which $q$ takes its maximum.  At that point, we set $q =
q_{\rm max}$, and at the other two points, which lie on the surface,
we have $q = 0$.  The potential $\phi_{\rm NS}$ is interpolated to
these three points, yielding $\phi_{\rm NS}^A$, $\phi_{\rm NS}^B$ and
$\phi_{\rm NS}^C$.  As we have discussed above, the neutron star
potential $\phi_{\rm NS}$ depends implicitly on $\bar r_e$.  To make
this dependence explicit, we compute $\hat \phi_{\rm NS} = \phi_{\rm
NS}/\bar r_e^2$ at the three points, which finally results in the
following three equations
\begin{eqnarray}
C & = & \bar r_e^2 \hat \phi_{\rm NS}^A 
- \frac{\bar M_{\rm BH}}{\bar r_e \hat r_{\rm BH}^A}
- \frac{1}{2} \hat \Omega^2 
(\hat x_A - \hat x_{\rm rot})^2 \nonumber \\
C & = & \bar r_e^2 \hat \phi_{\rm NS}^B 
- \frac{\bar M_{\rm BH}}{\bar r_e \hat r_{\rm BH}^B}
- \frac{1}{2} \hat \Omega^2 
(\hat x_B - \hat x_{\rm rot})^2 \\
C & = & \bar r_e^2 \hat \phi_{\rm NS}^C 
- \frac{\bar M_{\rm BH}}{\bar r_e \hat r_{\rm BH}^C}
- \frac{1}{2} \hat \Omega^2 
(\hat x_C - \hat x_{\rm rot})^2 + (n + 1) q_{\rm max}. \nonumber 
\end{eqnarray}
These equations can be solved iteratively for the eigenvalues $\hat
\Omega$, $\bar r_e$ and $C$.  Without the assumption of extreme mass
ratios a fourth condition has to be added so that $\hat x_{\rm rot}$
can be determined together with the three other eigenvalues.

\subsubsection{Rescaling of neutron star potential}

With the new value of $\bar r_e$, the rescaled neutron star potential
is computed from 
\begin{equation}
\phi_{\rm NS} = \bar r_e^2 \hat \phi_{\rm NS}.
\end{equation}

\subsubsection{Solution of the integrated Euler equation}

Given the three constants $\hat \Omega$, $\bar r_e$ and $C$ the
integrated Euler equation (\ref{bern4}) can now be solved everywhere.  The
innermost closed surface centered on the origin on which the density
$q$ equals zero is identified as the stellar surface.  Immediately
outside this surface, $q$ derived from equation (\ref{bern4}) becomes
negative; once we identify the surface, we properly set $q$ to zero
everywhere outside.  After obtaining the solution of the integrated Euler
equation, we evaluate the residual of the Poisson equation.  If this
residual is smaller than a specified tolerance, the iteration is
terminated; otherwise another full iteration is performed.

This completes the construction of a Newtonian BHNS binary for a given
black hole mass $\bar M_{\rm BH}$, a binary separation $\hat x_{\rm
BH}$ and a maximum neutron star density $q_{\rm max}$.  Binaries with
neutron stars of a given mass can then be constructed iteratively by
varying $q_{\rm max}$, and constant mass sequences can be computed by
varying $\hat x_{\rm BH}$.

%
%

\section{BHNS Binaries in General Relativity}
\label{sec3}

Constructing fully relativstic BHNS binaries amounts to finding a
spacetime metric
\begin{equation} \label{metric}
ds^2 = g_{ab} dx^a dx^b = 
- \alpha^2 dt^2 + \gamma_{ij} (dx^i + \beta^i dt)(dx^j + \beta^j dt)
\end{equation}
that solves Einstein's equations together with a matter distribution
\begin{equation}
T_{ab} = (\rho_0 + \rho_i + P) u_a u_b + P g_{ab}
\end{equation}
that satisfies the relativistic equations of hydrodynamics.  In the
above ``ADM'' \cite{adm62} form of the metric, $\alpha$ is the lapse,
$\beta^i$ is the shift vector, and $\gamma_{ij}$ is the spatial metric. 
In the stress-energy tensor, $\rho_i$ is the internal energy density
and $u^a$ the fluid four-velocity.

In this paper we focus on the construction of quasiequilibrium initial
data, so that we only seek to find a solution to the constraint
equations of Einsteins equations.  We solve these constraints within
the conformal thin-sandwich decomposition, which provides a natural
framework for the construction of quasiequilibrium (Section
\ref{thin-sandwich}, compare \cite{y99,c00,bs03}).  As in Newtonian
gravitation, the assumption of equilibrium allows us to model the star as
a solution to an integrated Euler equation (Section \ref{Bernoulli}).  The
numerical implementation of these equations again requires a
rescaling, which we discuss in Sections (\ref{dim}) and (\ref{scale}).

The constraint equations fix only some of the gravitational fields,
while others, which determine the ``background'' geometry, can be
chosen freely, but according to the astrophysical context.  In our
approach to constructing BHNS binaries, we choose for the background a
Schwarzschild black hole in Kerr-Schild coordinates (Section
\ref{kerr-schild}).  Our background therefore describes the black hole
in the binary and plays a role that is equivalent to the analytic
black hole potential $\phi_{\rm BH}$ in the Newtonian context.  We
then solve the constraints together with the integrated Euler equation to
model a neutron star in the presence of this black hole.

\subsection{The conformal thin-sandwich equations}
\label{thin-sandwich}

The conformal thin-sandwich decomposition has been used extensively
for the construction of binary neutron stars as well as binary black
hole systems (e.g.~\cite{wm95,bcsst,ggb02}), and has been developed
independently in \cite{y99}.  We refer to the reviews \cite{c00,bs03}
for a detailed derivation, and only state the most important results 
here.

The spatial metric $\gamma_{ij}$ is conformally decomposed into a
conformally related background metric $\tilde \gamma_{ij}$ and a
conformal factor $\psi$,
\begin{equation}
\gamma_{ij} = \psi^{4} \tilde \gamma_{ij}.
\end{equation}
Below we will assume that $\tilde \gamma_{ij}$ is given by a
Kerr-Schild metric.  For the construction of quasiequilibrium data in
a corotating frame it is natural to assume that the time derivative of
the conformally related metric vanishes,
\begin{equation}
\tilde u^{ij} \equiv \partial_t \tilde \gamma_{ij} = 0.
\end{equation}
For the trace of the extrinsic curvature we choose that of the
Kerr-Schild background, and set its time derivative to zero.  With
this choice all the freely specifiable quantities are fixed, and
Einstein's equations can be used to determine the lapse $\alpha$, the
shift $\beta^i$, and the conformal factor $\psi$.

The trace-free part of the evolution equation for the spatial metric
yields
\begin{equation} \label{Aij}
\tilde A^{ij} = \frac{\psi^6}{2 \alpha} (\tilde L \beta)^{ij},
\end{equation}
where $A^{ij}$ is the trace-free part of the extrinsic curvature
$K^{ij}$, $\tilde A^{ij} = \psi^{10} A^{ij}$ is the conformally
related trace-free extrinsic curvature, and where we have used $\tilde
u_{ij} = 0$.  The longitudinal operator $\tilde L$ is defined as
\begin{equation}
(\tilde L \beta)^{ij} \equiv \tilde D^i \beta^j + \tilde D^j \beta^i - 
	\frac{2}{3} \tilde \gamma^{ij} \tilde D_k \beta^k.
\end{equation}
Equation (\ref{Aij}) can now be inserted into the momentum constraint,
which yields an equation for the shift $\beta^i$
\begin{equation} \label{Mom}
\tilde \Delta_L \beta^i - 
(\tilde L \beta)^{ij} \tilde D_j \ln (\alpha \psi^{-6})
- \frac{4}{3} \alpha \tilde D^i K = 16 \pi \alpha \psi^4 \jmath^i.
\end{equation}
Here $j^a = - \gamma^a_{~b} n_c T^{bc}$ is the momentum density
observed by a normal observer and the vector Laplacian $\tilde
\Delta_L$ is
\begin{equation} \label{veclap}
\tilde \Delta_L \beta^i \equiv \tilde D_j (\tilde L \beta)^{ij} =
\tilde D^2 \beta^i + \frac{1}{3} \tilde D^i \tilde D_j \beta^j
+ \tilde R^i_{~j} \beta^j,
\end{equation}
where $\tilde R^i_{~j}$ is the Ricci tensor associated with the 
conformally related metric $\tilde \gamma_{ij}$.

The conformal factor $\psi$ is then determined from the Hamiltonian
constraint
\begin{equation} \label{Ham}
\tilde D^2 \psi - \frac{1}{8} \psi \tilde R
- \frac{1}{12} \psi^5 K^2 + \frac{1}{8} \psi^{-7} \tilde A_{ij} \tilde A^{ij}
= - 2 \pi \psi^5 \rho,
\end{equation}
where $\rho = n_a n_b T^{ab}$ is the energy density observed by a normal
observer and where
\begin{equation}
\tilde D^2 \psi \equiv \tilde \gamma^{ij} \tilde D_i \tilde D_j \psi
	=  \tilde \gamma^{ij} \psi_{,ij} - \tilde \Gamma^i \psi_{,i}
\end{equation}
is the Laplace operator associated with $\tilde \gamma_{ij}$.

To derive an equation for the lapse we assume 
\begin{equation}
\partial_t K = 0
\end{equation}
(but note that $K \ne 0$).  Using the trace of the evolution equation
for the extrinsic curvature we find that this condition implies
\begin{eqnarray} \label{Kdot}
D^2 \alpha & = & \alpha (K_{ij} K^{ij} + 4 \pi (\rho + S)) + \beta^i D_i K \\
	& = &  \alpha (\psi^{-12} \tilde A_{ij} \tilde A^{ij} + \frac{1}{3} K^2
	 + 4 \pi (\rho + S)) + \beta^i D_i K ,\nonumber
\end{eqnarray}
where $S = \gamma^{ab} T_{ab}$.  This equation involves the
Laplace operator with respect to the physical metric $\gamma_{ij}$,
which we do not know a priori.  It is therefore convenient to rewrite
this operator in terms of the conformal background metric $\tilde
\gamma_{ij}$.  To do so, we combine equation (\ref{Kdot}) with the
Hamiltonian constraint (\ref{Ham}), which yields
\begin{eqnarray} \label{lapse}
\tilde D^2 (\alpha \psi) & =  &
	\frac{7}{8} \alpha \psi^{-7} \tilde A_{ij} \tilde A^{ij} 
	+ \frac{5}{12} \alpha \psi^5 K^2
	+ \frac{1}{8} \alpha \psi \tilde R \nonumber \\
& &	+ \psi^5 \beta^i \tilde D_i K
	+ 2 \pi \alpha \psi^5 (\rho + 2 S).
\end{eqnarray}
Equation (\ref{lapse}) is the generalization of the $\partial_t K = 0$
condition for a non-zero (but constant in time) $K$. It reduces to the
more familiar maximal slicing equation for $K = 0$.

Equations (\ref{Ham}), (\ref{Mom}) and (\ref{lapse}) now form the
conformal thin-sandwich equations for the unknown 
gravitational field quantities $\psi$, $\beta^i$ and $\alpha$.  These
equations are equivalent to Poisson's equation (\ref{poisson1})
in the Newtonian framework.

\subsection{The integrated Euler equation}
\label{Bernoulli}

For corotating equilibrium solutions, the relativistic Euler
equation can be integrated analytically to yield
\begin{equation} \label{bern_rel}
\frac{h}{u^t} = 1 + C,
\end{equation}
where $C$ is a constant (we add one to the right hand side so that $C$
reduces to the equivalent constant in the integrated Newtonian Euler
equation (\ref{bern1})), and where $h$ is the specific enthalpy
\begin{equation} \label{enth1}
h = \exp{\int \frac{dP}{\rho_0 + \rho_i + P}}
\end{equation}
(see, e.g., \cite{bs03}.) Here $\rho_i$ is the internal energy
density.  Note that the integrated Euler equation is consistent with
von Zeipel's theorem, which states that in uniformly rotating, perfect
fluid stars, surfaces of contant $P$, $\rho_0$ and $u^t$ all coinide.

The time component of the four-velocity $u^t$ can be expressed in
terms of the relative velocity $v$ between the matter and a normal
observer as
\begin{equation} \label{ut}
\alpha u^t = \frac{1}{(1 - v^2)^{1/2}}.
\end{equation}
In a corotating coordinate system, where $u^i = 0$, $u^t$ can be 
found from the normalization condition $u_a u^a = -1$, which 
yields
\begin{equation} \label{ut2}
u^t = \left( \alpha^2 - \gamma_{ij} \beta^i \beta^j \right)^{-1/2}.
\end{equation}
This expression, together with the enthalpy (\ref{enth1}), can be 
inserted into the integrated Euler equation (\ref{bern_rel}), yielding
a relation between the fluid and gravitational field variables.

As before, we adopt a polytropic equation of state (\ref{eos}) and
introduce the density parameter $q$ (\ref{q}).  The enthalpy
(\ref{enth1}) can then be written as
\begin{equation} \label{enth2}
h = \frac{\rho_0 + \rho_i + P}{\rho_0} = 1 + (1 + n) q,
\end{equation}
where we have used
\begin{equation}
\rho_i = nP = n \kappa^{-n} q^{n+1}.
\end{equation}
The integrated Euler equation then becomes
\begin{equation} \label{bern5}
q = \frac{1}{1+n} \left( u^t (1+C) - 1 \right).
\end{equation}
In the Newtonian limit this equation reduces to the integrated Newtonian
Euler equation (\ref{bern2a}).

\subsection{Elimination of Dimensions}
\label{dim}

As in Section (\ref{Nelimdim}) for the Newtonian treatment, we can now
eliminate dimensions by rescaling all dimensional quantities with
respect to appropriate powers of $\kappa^{n/2}$, which has units of
length.  In particular we have $\bar D = \kappa^{n/2} \tilde D$, $\bar
R = \kappa^n \tilde R$, $\bar K = \kappa^{n/2} K$, and $\bar A^{ij} =
\kappa^{n/2} \tilde A^{ij}$.  Denoting these dimensionless quantites
with bars, we arrive at the following system of field equations
\begin{eqnarray}
\bar A^{ij} & = & \frac{\psi^6}{2 \alpha} (\bar L \beta)^{ij}  \nonumber \\
\bar D^2 \psi & = &
\frac{1}{8} \psi \bar R
+ \frac{1}{12} \psi^5 \bar K^2 - \frac{1}{8} \psi^{-7} \bar A_{ij} \bar A^{ij}
- 2 \pi \psi^5 \bar \rho  \nonumber \\
\bar \Delta_L \beta^i & = &
(\bar L \beta)^{ij} \bar D_j \ln (\alpha \psi^{-6})
+ \frac{4}{3} \alpha \bar D^i \bar K + 16 \pi \alpha \psi^4 \bar \jmath^i  
\nonumber\\
\bar D^2 (\alpha \psi) & =  &
	\frac{7}{8} \alpha \psi^{-7} \bar A_{ij} \bar A^{ij} 
	+ \frac{5}{12} \alpha \psi^5 \bar K^2
	+ \frac{1}{8} \alpha \psi \bar R \nonumber \\
& &	+ \psi^5 \beta^i \bar D_i \bar K
	+ 2 \pi \alpha \psi^5 (\bar \rho + 2 \bar S). \label{field1}
\end{eqnarray}
Here the matter sources can be expressed in terms of the dimensionless
quantity $q$ 
\begin{eqnarray}
\bar \rho & = & \kappa^{n} \rho =  
q^n \left(\frac{1 + (1+n)q}{1 - v^2} - q \right) 
\nonumber \\
\bar \jmath^i & = & \kappa^{n} j = 
q^n \frac{1 + (1+n)q}{1 - v^2} \, \frac{\beta^i}{\alpha}
\label{matter1} \\
\bar S & = & \kappa^{n} S = 
q^n \left(\frac{1 + (1+n)q}{1 - v^2}v^2 + 3 q \right).
\nonumber
\end{eqnarray}
The field equations (\ref{field1}) have to be solved together with
the integrated Euler equation (\ref{bern5}).

\subsection{Rescaling}
\label{scale}

As in the Newtonian problem, we would like the star's surface to
intersect the $x$-axis at constant coordinate values of $\hat x_A =
-1$ and $\hat x_B = 1$.  This again requires a rescaling with respect
to the star's dimension, which we again call $\bar r_e$ (see Section
\ref{Nrescale}).  Denoting these rescaled quantities with hats,
e.g.~$\hat D = \bar r_e \bar D$, $\hat R = \bar r_e^2 \bar R$, $\hat
A^{ij} = \bar r_e \bar A^{ij}$, $\hat K = \bar r_e \bar K$, we find
the Hamiltonian constraint
\begin{equation} \label{Ham1}
\hat D^2 \psi =
\frac{1}{8} \psi \hat R
+ \frac{1}{12} \psi^5 \hat K^2 - \frac{1}{8} \psi^{-7} \hat A_{ij} \hat A^{ij}
- 2 \pi \psi^5 \bar r_e^2 \bar \rho,
\end{equation}
the Momentum constraint
\begin{equation} \label{Mom1}
\hat \Delta_L \beta^i = 
(\hat L \beta)^{ij} \hat D_j \ln (\alpha \psi^{-6})
+ \frac{4}{3} \alpha \hat D^i \hat K 
+ 16 \pi \alpha \psi^4 \bar r_e^2 \bar \jmath^i,  
\end{equation}
and the lapse equation
\begin{eqnarray} \label{lapse1}
\hat D^2 (\alpha \psi) & =  &
	\frac{7}{8} \alpha \psi^{-7} \hat A_{ij} \hat A^{ij} 
	+ \frac{5}{12} \alpha \psi^5 \hat K^2
	+ \frac{1}{8} \alpha \psi \hat R \nonumber \\
& &	+ \psi^5 \beta^i \hat D_i \hat K
	+ 2 \pi \alpha \psi^5 \bar r_e^2 (\bar \rho + 2 \bar S). 
\end{eqnarray}
In the above equations, the extrinsic curvature is computed from
\begin{equation} \label{extcurv}
\hat A^{ij} = \frac{\psi^6}{2 \alpha} (\hat L \beta)^{ij}.
\end{equation}
Inserting (\ref{ut2}) into (\ref{bern5}) we also find the 
integrated Euler equation
\begin{equation} \label{bern6}
q = \frac{1}{1+n} \left( \frac{1 + C}{(\alpha^2 - 
\psi^4 \hat \gamma_{ij} \beta^i \beta^j)^{1/2}} - 1 \right).
\end{equation}
We note that the metric is dimensionless, so that
$\hat \gamma_{ij} = \bar \gamma_{ij} = \tilde \gamma_{ij}$.  For ease
of notation, we decorate the conformally related metric with a hat.

\subsection{Kerr-Schild Background}
\label{kerr-schild}

As discussed above, we account for the presence of the black hole by
choosing a background metric that, in the absence of the neutron star,
describes a Schwarzschild black hole.  This black hole can be
expressed in different coordinate systems.  Probably the simplest
choice would be standard isotropic coordinates (in which case the
presence of the black hole is actually absorbed in the conformal
factor, while the background 3-metric is flat).  The disadvantage of
these coordinates is that they do not penetrate into the black hole
interior.  We expect that such penetration is crucial when dealing
with companions of comparable mass, in which case the spacetime in the
vicinity of the black hole and its apparent horizon have to be solved
for.  Painlev\'e-Gullstrand coordinates are both remarkable simple and
regular on the black hole horizon (see \cite{mp01} for a recent
discussion).  However, these coordinates lead to spatial slices that
are not asymptotically flat, complicating the definition of mass and
angular momentum.  We have therefore chosen to adopt Kerr-Schild (or
ingoing Eddington-Finkelstein coordinates), in terms of which the
Schwarz\-schild metric for a single, nonrotating black hole can be
written as
\begin{equation} \label{KSmetric}
ds^2 = g_{ab} dx^a dx^b = (\eta_{ab} + 2 H l_a l_b) dx^a dx^b,
\end{equation}
(compare \cite{c92,mhs99,mm00,pct02}.)  Here
\begin{equation}
H = \frac{M_{\rm BH}}{r_{\rm BH}} = 
\frac{\bar M_{\rm BH}}{\bar r_{\rm BH}} = 
\frac{1}{\bar r_e} \, \frac{\bar M_{\rm BH}}{\hat r_{\rm BH}},
\end{equation}
is the analogue of the Newtonian potential (\ref{scale_phi_BH1}) and
scales in exactly the same way.  As in the Newtonian case 
\begin{equation}
\hat r_{\rm BH} = \left [ (\hat x - \hat x_{\rm BH})^2
                        + (\hat y - \hat y_{\rm BH})^2
                        + (\hat z - \hat z_{\rm BH})^2 \right]^{1/2}
\end{equation}
is the coordinate distance to the center of the black hole at the
coordinate location $(-\hat x_{\rm BH},0,0)$ ($r_{\rm BH}$ is the
familiar areal radial coordinate centered on the Schwarzschild black
hole).  The vector $l^a$ is null with respect to both $g_{ab}$ and the
``Minkowski'' metric $\eta_{ab}$ and can be written as
\begin{equation}
l_t = - l^t = 1,~~~~~~~~~~~l_i = l^i = \frac{x^i}{r_{\rm BH}}.
\end{equation}

Comparing the Kerr-Schild metric (\ref{KSmetric}) with the ADM form of
the metric (\ref{metric}) shows that we can identify the lapse as
\begin{equation} \label{KSlapse}
\alpha_{\rm BH} = (1 + 2 H)^{-1/2},
\end{equation}
the shift as 
\begin{equation} \label{KSshift}
\beta^i_{\rm BH} = 2 H \alpha^2_{\rm BH} l^i,
\end{equation}
and the spatial metric as
\begin{equation}
\hat \gamma_{ij} = \eta_{ij} + 2 H l_i l_j. 
\end{equation}
Here we use the subscript BH to distinguish these background
quantities from the lapse $\alpha$ and shift $\beta^i$ of the full
solution.  

We note that $l_a$ is a four-vector, which is therefore raised with the
spacetime metric
\begin{equation}
l^{a} = (\eta^{ab} - 2 H l^a l^b) l_b = \eta^{ab} l_b
\end{equation}
as opposed to the spatial metric.  The inverse of the spatial metric
is therefore given by
\begin{equation} \label{KSsm}
\hat \gamma^{ij} = \alpha^2_{\rm BH} 
\left( ( 1 + 2 H ) \eta^{ij} - 2 H l^i l^j \right). 
\end{equation}
Other useful quantities that appear in many of the equations are
\begin{eqnarray}
\hat \Gamma^i_{jk} & = & 
\frac{\alpha_{\rm BH}^2 H}{\hat r_{\rm BH}} 
( 2 l^i \eta_{jk} - 3 l^i l_j l_k) \\
\hat \Gamma^i & \equiv & \hat \gamma^{jk} \hat \Gamma^i_{jk} = 
	\frac{\alpha_{\rm BH}^4 H}{\hat r_{\rm BH}} 
	(3 + 8 H) l^i \label{KScf} \\
\hat K & = & \frac{2 \alpha_{\rm BH}^3 H}{\hat r_{\rm BH}} (1 + 3 H).
\end{eqnarray}
Several other quantities will be required for the solution of the
thin-sandwich equations in a Kerr-Schild background, and we will
provide those as they appear in the equations.

Just like the Newtonian black hole potential (\ref{NphiBH}) is a
vaccum solution to the Poisson equation (\ref{poisson1}), the
background lapse (\ref{KSlapse}), shift (\ref{KSshift}) and the
trivial background conformal factor $\psi_{\rm BH} = 1$ solve the
field equations (\ref{Ham1}) -- (\ref{lapse1}) identically in the
absence of matter sources, $\bar \rho = \bar \jmath^i = \bar S = 0$.
In further analogy with the Newtonian problem we therefore solve the
relativistic equations by computing corrections to the background
quantities that account for the presence of the neutron star.  Details
of the numerical strategy and implementation are presented in Appendix
\ref{appA}.

Once a model has been constructed, we can compute the neutron star's
rest mass from
\begin{equation}
M_0 = \int \rho_0 u^{\alpha} d^3 \Sigma_\alpha = 
\int \rho_0 u^t \sqrt{-g} d^3 x.
\end{equation}
In terms of the variables used in our code this can be rewritten as
\begin{equation} \label{m_0}
\bar M_0 \equiv \kappa^{-n/2} M_0 = 
\bar r_e^3 \int \alpha \psi^6 u^t \sqrt{\gamma_{\rm BH}} q^n d^3 \hat x,
\end{equation}
where $\gamma_{\rm BH}$ is the determinant of the spatial Kerr-Schild
background metric.

%
%

\section{Extreme Mass Ratios}
\label{sec4}

In this paper we focus on the construction of neutron stars in the 
presence of a black hole companion by assuming extreme mass ratios
$\bar M_{\rm BH} >> \bar M_{\rm NS}$.  This approximation simplifies
the problem in a number of ways.

Most importantly, we may assume that the spacetime is affected by
the presence of the neutron star only in a neighborhood of the
neutron star.  The numerical grid therefore needs to cover
only a region around the neutron star, and does not need to encompass
the black hole.  That means that we do not need to excise the black
hole interior, simplifying the computational domain and eliminating
the need for interior black hole boundary conditions.  This also means
that the computational grid can be chosen significantly smaller, 
which speeds up numerical calculations.

Another simplification arises from the fact that in the
extreme-mass-ratio limit, the binary's center of mass resides at the
center of the black hole, $\hat x_{\rm rot} = \hat x_{\rm BH}$, which
eliminates the need to locate $\hat x_{\rm rot}$ iteratively.  In the
general case, the radius $\hat x_{\rm rot}$ can be identified, for
example, by noting that the solution becomes quasistationary in the
corotating frame and therefore must satisfy the additional condition
that $M_{\rm ADM} = M_{\rm Komar}$ when the center of mass is
correctly located \cite{ggb02} (see \cite{sb02} for an illustration).
Finally, for the construction of constant-mass sequences, the
irreducible mass of the black hole has to be held constant, which
requires an additional iteration.  This iteration is also unnecessary
in the extreme mass ratio limit.

%
%

\section{Numerical Results}
\label{sec5}

\begin{figure}
\includegraphics[scale=0.4]{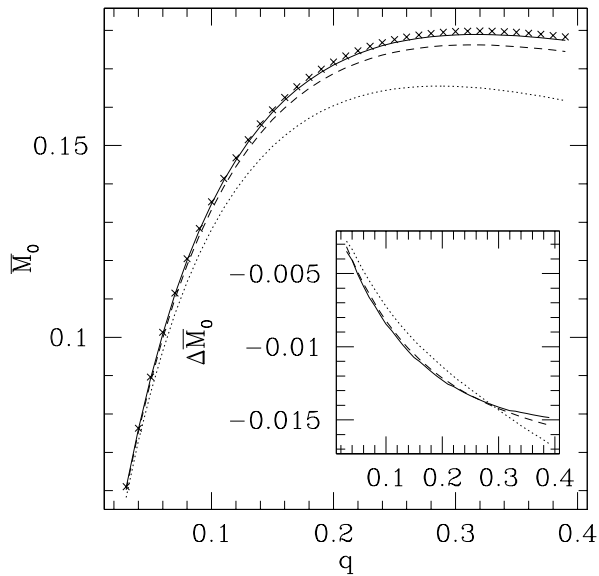} 
\caption{Rest mass $\bar M_0$ of $n=1$ polytropes in isolation as
function of the density parameter $q$.  The crosses denote the exact
results for spherical, non-rotating neutron stars, and the lines
correspond to numerical results for a fine grid (16 gridpoints across
a neutron star radius; denoted by the solid line), a medium grid (8
points; dashed line) and a coarse grid (4 points; dotted line).  In
the inset we show the rescaled numerical error, i.e.~the difference
between the fine-grid solution and the exact solution, 
multiplied by a factor of 16 and the difference between the medium grid 
solution and exact solution, multiplied by factor of 4.  
The convergence of the rescaled error functions
establishes second-order convergence of our implementation.}
\label{TOV_M}
\end{figure}

\begin{figure}
\includegraphics[scale=0.4]{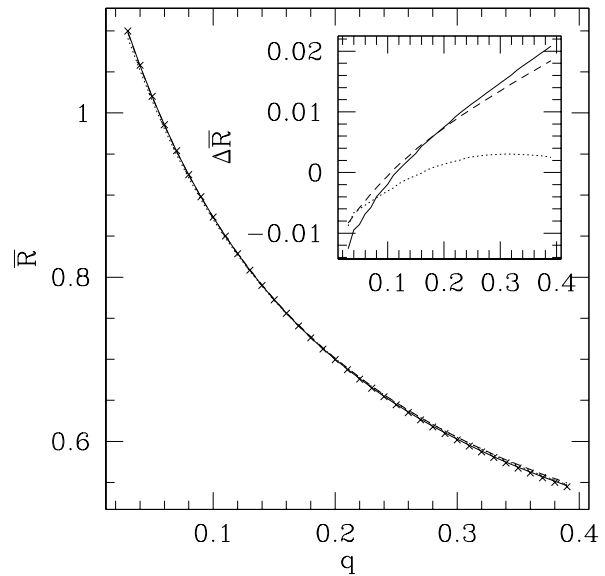} 
\caption{Same as Fig.~\ref{TOV_M}, except for the isotropic neutron
star radius $\bar R$.  The three different resolutions are almost
indistinguishable and agree well with the exact solution marked by the
crosses.  However, the inset with the rescaled numerical errors shows
that the coarse resolution with only 4 grid points across the neutron
star radius is not yet in the convergent regime.}
\label{TOV_R}
\end{figure}

In this Section we present numerical results, both for neutron stars
in isolation (as a code check, Section \ref{TOV}) and for binary
sequences of variable separation but constant rest mass (Section
\ref{sequences}).  We discuss tidal disruption in Section \ref{tidal}.
For all results in this Section we adopt a polytropic index of $n =
1$.

\subsection{Neutron Stars in Isolation}
\label{TOV}

As a simple code test we can use our formalism to construct neutron
stars in isolation ($\bar M_{\rm BH} = 0$), for which the solution is
given by the non-rotating, spherical Tolman-Oppenheimer-Volkoff (TOV)
solutions \cite{ov39}.  TOV solutions can be considered ``exact''
since they satisfy a set of ordinary differential equations that
essentially can be solved to arbitrary accuracy.

In Figs.~\ref{TOV_M} and \ref{TOV_R} we show the exact values of the
rest mass $\bar M_0$ and the {\it isotropic} radius $\bar R$ as a
function of the density parameter $q$, together with our numerical
results for three different grid resolutions.  For these tests we
imposed the outer boundaries at $\hat X_{\rm out}$ = 2 (see Section
\ref{num_grid}), i.e.~at twice the neutron star radius (see Sections
\ref{Nrescale} and \ref{scale}).  We adopted numerical grids of $(64
\times 64 \times 32)$, $(32 \times 32 \times 16)$ and $(16 \times 16
\times 8)$, corresponding to 16, 8 and 4 gridpoints along a neutron
star radius.  Focussing on the rest mass $\bar M_0$ plot, it can be
seen how the numerical results converge to the exact results.  In the
insets we also show the rescaled numerical errors, which establish
second order convergence of our code.  We find very similar results
for the gravitational (ADM) mass $\bar M$.

\subsection{Constant Rest Mass Sequences}
\label{sequences}

\begin{figure}
\includegraphics[scale=0.4]{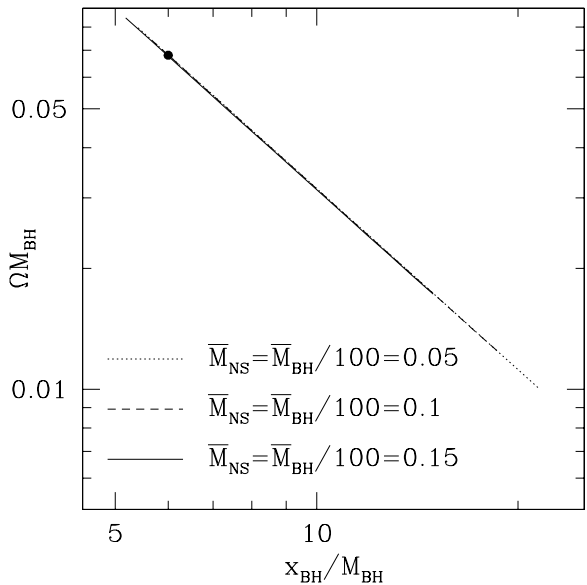} 
\caption{The orbital frequency as a function of binary separation for
mass ratios $M_{\rm BH}/M_{\rm NS} = 100$ and three different neutron
star rest masses $\bar M_{\rm NS} = 0.05$, 0.1 and 0.15.  Our results
typically agree with the Kepler frequency (\ref{kepler}) to within one
percent.  The dot marks the Kepler frequency $\Omega M_{\rm BH} =
6^{-3/2}$ at the innermost stable circular orbit at $x_{\rm BH} = 6
M_{\rm BH}$.  For these simulations we used a grid of $(64 \times 64
\times 32)$ gridpoints and imposed outer boundaries at $\hat X_{\rm
out}$ = 4.}
\label{Omega}
\end{figure}

Inspiraling BHNS binaries conserve the rest mass of the neutron star
and the irreducible mass of the black hole.  We construct sequences of
constant neutron star rest mass by iterating, for a given binary
separation $x_{\rm BH}$, over the maximum density parameter $q_{\rm
max}$ until a neutron star of the desired rest mass has been found to
within a specified tolerance.  

Keeping the black hole's irreducible mass constant would involve
evaluating the area of its event horizon (or, in practice, its
apparent horizon).  The difference between the black hole background
mass $\bar M_{\rm BH}$ and the black hole's irreducible mass is in the
order of the binary's binding energy \cite{fn1}.  In this paper we
neglect effects of the neutron star on the black hole, and hence
neglect this change in the black hole mass.  Strictly speaking, we
therefore construct sequences of constant neutron star rest mass $\bar
M_0$ and black hole background mass $\bar M_{\rm BH}$.

In the test-mass limit, the orbital frequency $\Omega$ is given by
the relativistic version Kepler's third law
\begin{equation} \label{kepler}
\Omega M_{\rm BH} = \left( \frac{x_{\rm BH}}{M_{\rm BH}} \right)^{-3/2}.
\end{equation}
In Fig.~\ref{Omega} we plot $\Omega M_{\rm BH}$ for mass ratios
$M_{\rm BH}/M_{\rm NS} = 100$ and three different neutron star rest
masses $\bar M_{\rm NS} = 0.05$, 0.1 and 0.15 (the maximum allowed
rest mass for isolated, $n=1$ polytropes in these units is $\bar
M_{\rm NS} = 0.18$.)  The corresponding compaction $M_{NS}/R_{\rm
areal}$ of these neutron stars in isolation are 0.042, 0.088 and 0.145
(here $R_{\rm areal}$ is the neutron star's areal radius).  All
frequencies typically agree with the relativistic Kepler frequency to
within better than one percent, and, as expected, the frequency is
independent of the neutron star mass.  The dot in Fig.~\ref{Omega}
marks the orbital frequency $\Omega M_{\rm BH} = 6^{-3/2} = 6.804$ at
the innermost stable circular orbit at $x_{\rm BH} = 6 M_{\rm BH}$.

\subsection{Tidal Disruption}
\label{tidal}

\begin{figure}
\includegraphics[scale=0.4]{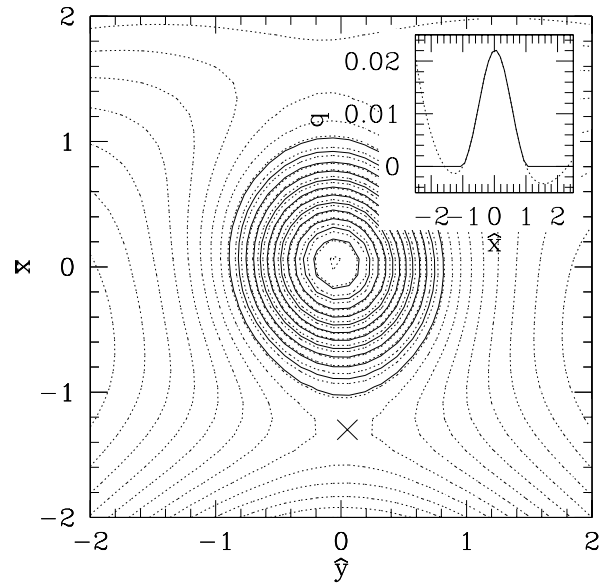} 
\caption{Contours of the density parameter $q$ (solid lines) and the
right hand side of the integrated Euler equation (\ref{bern5}) (dotted lines)
for $\bar M_{\rm NS} = \bar M_{\rm BH}/10 = 0.05$ at a separation of
$\bar x_{\rm BH} = 12.6 \bar M_{\rm BH}$ (the center of the black hole
is located at $\hat x = - 5.0$ and $\hat y = 0$).  The innermost
contour has a density ratio $q/q_{\rm max} = 10/11$, and successive
contours have q decreasing by the same decrements $\Delta q/q_{\rm
max} = 1/11$.  The saddle point between the neutron star and the black
hole defines the inner Lagrange point $L_1$ and is marked by the
cross.  The equipotential surface passing through $L_1$ is the
relativistic Roche lobe.  The proximity of $L_1$ to the neutron star
surface indicates that the neutron star is close to being tidally
disrupted.  The saddle point ``above'' the neutron star is the outer
Lagrange point $L_2$.  The inset shows both $q$ (solid line) and the
right hand side of (\ref{bern5}) (dotted line) along the $x$-axis as a
function of $\hat x$.  The saddle point is close to the minimum of the
dotted line at about $\hat x = - 1.3$. }
\label{contour1}
\end{figure}

\begin{figure}
\includegraphics[scale=0.4]{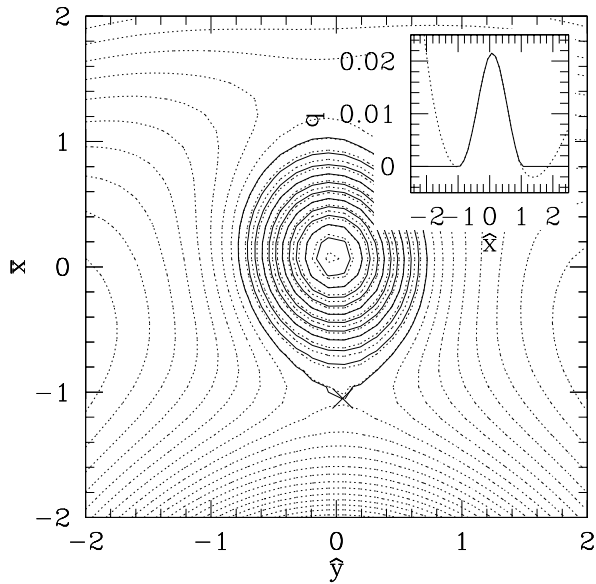} 
\caption{Same as Fig.~\ref{contour1}, except for a separation of $\bar
x_{\rm BH} = 11.9 \bar M_{\rm BH}$ (the center of the black hole is
located at $\hat x = - 4.18$ and $\hat y = 0$).  The Lagrange point
$L_1$ is now located on the star's surface, indicating the onset of
tidal disruption.}
\label{contour2}
\end{figure}

\begin{table}
\begin{center}
\begin{tabular}{ccccc}
\tableline \tableline 
$\hat x_{\rm BH}$ & $x_{\rm BH}/M_{\rm BH}$ & $\bar r_e$ &  $q_{\rm max}$ &
$\Omega M_{\rm BH}$ \\
\tableline 
- 8.0   &  17.8  & 1.11   & 0.0235  & 0.0133 \\
- 5.0	&  12.6  & 1.26   & 0.0223  & 0.0223 \\
- 4.74  &  12.3  & 1.30   & 0.0221  & 0.0230 \\
- 4.49 	&  12.1  & 1.35   & 0.0220  & 0.0236 \\
- 4.34  &  12.0  & 1.38   & 0.0219  & 0.0239 \\
- 4.18  &  11.9  & 1.42   & 0.0218  & 0.0241 \\
\tableline \tableline 
 \end{tabular}
\end{center}
\caption{Numerical values for a constant rest-mass sequence with $\bar
M_{\rm BH} = 0.5$ and $\bar M_{\rm NS} = 0.05$ close to the onset of
tidal disruption at a separation of $x_{\rm BH} = 11.9 M_{\rm BH}$
(see Fig.~\ref{contour2}).}
\label{Seq_table}
\end{table}

In corotating binaries, tidal disruption occurs when the neutron star
overflows its Roche lobe.  In Newtonian physics, the Roche lobe is
defined as the innermost equipotential surface of the effective
potential (\ref{phi_eff}) that encompasses both binary companions.  In
a contour plot in the equatorial plane the Roche lobe forms a ``figure
eight'', in which each teardrop-shaped region contains one companion.
The equipotenial surface pinches off at the saddle-point of the
potential, which is called the inner Lagrange point $L_1$.  When the
neutron star overflows its Roche lobe, matter flows across this inner
Lagrange point and accrets onto the black hole.

By virtue of the integrated Euler equation (\ref{bern2a}) the density
parameter $q$ is proportional, up to a constant, to the effective
potential $\Phi_{\rm eff}$.  The Roche lobe can therefore be
constructed equivalently from contours of the right hand side of the
integrated Euler equation (\ref{bern2a}).  This immediately suggests
the construction of a relativistic Roche lobe from the right hand side
of the integrated relativistic Euler equation (\ref{bern5}).  For a
relativistic equilibrium configuration, the Roche lobe and Lagrange
points can therefore be identified in complete analogy to the
Newtonian case.

In Figs.~(\ref{contour1}) and (\ref{contour2}) we show contours for
two configurations with $\bar M_{\rm NS} = \bar M_{\rm BH}/10 = 0.05$
in the equatorial plane.  Table \ref{Seq_table} also provides
numerical values for this sequence.  Fig.~(\ref{contour1}) shows the
binary at a separation of $\bar x_{\rm BH} = 12.6 \bar M_{\rm BH}$
(the second line in Table \ref{Seq_table}).  For this calculation and
all others in this Section we used a grid of $(48 \times 48 \times
24)$ gridpoints and imposed outer boundaries at $\hat X_{\rm out}$ =
3.  Density contours are marked by solid lines, while contours of the
right hand side of (\ref{bern5}), i.e.~contours of the relativistic
effective potential, are marked by the dotted lines.  The inner
Lagrange point $L_1$ is marked by the cross.  The contour passing
through this point is the Roche lobe.  The outermost solid contour,
which marks the surface of the star, is still inside the Roche lobe,
so tidal disruption has not yet set in.

In Fig.~(\ref{contour2}) we show the same binary, but at a slightly
smaller separation of $\bar x_{\rm BH} = 11.9 \bar M_{\rm BH}$ (the
last line in Table \ref{Seq_table}).  Now the Lagrange point $L_1$,
again marked by a cross, is located on the surface of the star,
indicating that the star now fills out its Roche lobe, and that tidal
disruption sets in.  This illustrates how the onset of tidal
disruption can be identified by the Lagrange point $L_1$ reaching the
surface of the star, and simultaneously the star filling out its Roche
lobe.

\begin{figure}
\includegraphics[scale=0.4]{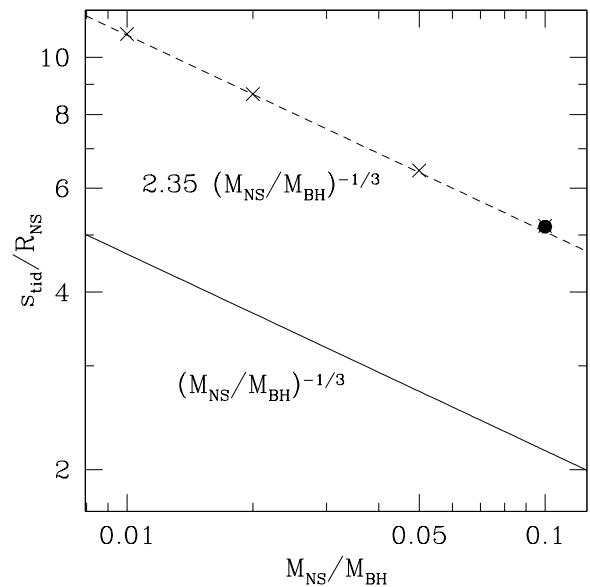} 
\caption{The tidal separation $s_{\rm tid}/R_{\rm NS}$ as a function
of mass ratio $M_{\rm NS}/M_{\rm BH}$.  The crosses mark our numerical
results for $\bar M_{\rm NS} = 0.01$, the dot marks a result obtained
by Ury\=u \& Eriguchi \cite{ue99}, and the solid line denotes the
expected qualitative scaling relation (\ref{m_ratio_scaling}),
$(M_{\rm NS}/M_{\rm BH})^{-1/3}$.  The dashed line, $2.35 (M_{\rm
NS}/M_{\rm BH})^{-1/3}$, provides an approximate fit to our numerical
data.}
\label{scaling}
\end{figure}

\begin{table}
\begin{center}
\begin{tabular}{cccc}
\tableline \tableline 
$M_{\rm NS}/M_{\rm BH}$  & $x_{\rm BH}/M_{\rm BH}$ & $s_{\rm tid}/R_{\rm NS}$ & $\Omega M_{\rm BH}$ \\
\tableline 
0.01	& 13.3 	& 11.0	&  0.0203 \\
0.02	& 20.9	& 8.67	&  0.0102 \\
0.05	& 38.5  & 6.43	&  0.00405 \\
0.1	& 61.7	& 5.18	&  0.00199 \\
\tableline
0.1	& ---	& 5.17  &  0.00208 \\
\tableline \tableline 
 \end{tabular}
\end{center}
\caption{Numerical values for the onset of tidal disruption of an
$\bar M_{\rm NS} = 0.01$ neutron star (whose isotropic radius in
isolation is $\bar R_{\rm NS} = 1.23$).  The bottom line are the
results from the Newtonian models of Ury\=u \& Eriguchi \cite{ue99}
(see their Table 2).}
\label{Tid_table}
\end{table}

It is now of interest to see whether we recover the scaling
(\ref{m_ratio_scaling}).  To better compare with Ury\=u \& Eriguchi
\cite{ue99}, who have located the tidal separation in Newtonian BHNS
binaries, we compute their measure of separation
\begin{equation}
s_{\rm tid} = x_{\rm BH} + \frac{1}{M_{\rm NS}} \int x \rho d^3 x
\end{equation}
(denoted $d_G$ in \cite{ue99}.)  In Table \ref{Tid_table} and
Fig.~\ref{scaling} we show results for $\bar M_{\rm NS} = 0.01$, for
which $R_{\rm NS}/M_{\rm NS} = 123$.  For these configurations
relativistic effects are sufficiently weak to make a comparison with
Newtonian results meaningful.  According to the scaling relation
(\ref{isco_scaling}), tidal disruption will occur over a reasonable
range of mass ratios $M_{\rm NS}/M_{\rm BH} \ll 1$.  Crosses in
Fig.~(\ref{scaling}) denote results from our code, and the dot marks
the only result from \cite{ue99} for $n=1$ and $M_{\rm NS}/M_{\rm BH}
\leq 0.1$ (see also the bottom line in Table \ref{Tid_table}).  Our
result agrees with that of \cite{ue99} to within a few percent (and
even better for the tidal separation $s_{\rm tid}$), which is well
within the error that one expects for a mass ratio of $M_{\rm
NS}/M_{\rm BH} = 0.1$ within our approximation $M_{\rm NS}/M_{\rm BH}
\ll 1$.  For this configuration both $M_{\rm NS}/R_{\rm NS} \approx
0.01$ and $M_{\rm BH}/s_{\rm tid} \approx 0.01$ so that relativistic
effects could also introduce a deviation of a few percent.

The scaling of the tidal separation with the mass ratio is very close
to that predicted by (\ref{m_ratio_scaling}); in fact, the crude
estimate for the tidal separation differs by only about a factor of
2.35 from our numerical results.

%
%

\section{Summary}
\label{sec6}

We have developed a new relativistic formalism for the construction of
BHNS binaries in quasiequilibrium.  We solve the constraint equations
of general relativity, decomposed in the thin-sandwich formalism,
together with the integrated Euler equation for the neutron star
matter.  The background geometry describes a Schwarzschild black hole,
which accounts for the neutron star's binary companion.  Our approach
yields self-consistent solutions describing a neutron star and its
self-gravity in the presence of a black hole companion.

In this paper we assume that the black hole mass is much larger than
that of the neutron star, which simplifies the problem in a number of
ways.  In this case, the axis of rotation, which in general is located
at the binary center of mass, passes through the center of the black
hole.  Also in this limit, we can assume that the neutron star only
affects the spacetime in a neighborhood around the star, so that the
numerical grid can be restricted to such a neighborhood and does not
need to cover the black hole.  This eliminates the need for black hole
excision.  Furthermore, it is appropriate to assume that the black
hole's irreducible mass, which is essentially constant during a
quasiequilibrium inspiral, is equal to the background mass, which
eliminates the need for an iteration.

Adopting this approximation, we have tested our code for isolated
neutron stars and for constant rest mass sequences, finding excellent
agreement with Kepler's third law for the orbital frequency.  We have
also developed a fully relativistic framework for the identification
of the Roche lobe and Lagrange points, which can be used to locate the
binary separation at which tidal break-up sets in.  For weakly
relativistic $n=1$ polytropes, for which the tidal separation is
outside of the innermost stable circular orbit even for large mass
ratios, we find that tidal disruption occurs at about twice the
separation suggested by the crude scaling relation $s_{\rm tid}/R_{\rm
NS} \sim (M_{\rm BH}/M_{\rm NS})^{1/3}$.  

Astrophysically more interesting scenarios require $M_{\rm BH} \sim
M_{\rm NS}$, for which the above assumption of extreme mass ratios has
to be relaxed. We plan to provide solutions to black hole -- neutron
star binaries with companions of comparable mass in the near future.

\acknowledgments

TWB would like to thank Carsten Gundlach, Eric Poisson and Charles
Evans for helpful discussions, and the Kavli Institute for Theoretical
Physics at the University of California Santa Barbara for its
hospitality.  MLS gratefully acknowledges support through a Goldwater
Fellowship and a Surdna Foundation Undergraduate Research Fellowship.
This paper was supported in part by NSF Grant PHY-0139907 at Bowdoin
College and NSF Grants PHY-0090310 and PHY-0345151 and NASA Grant NAG
5-10781 at the University of Illinois at Urbana-Champaign.

\begin{appendix}

%
%

\section{Numerical Strategy}
\label{appA}

In this Appendix we provide details on the numerical strategy and
implementation for the solution of the field equations (\ref{Ham1}) --
(\ref{lapse1}), together with the integrated Euler equation (\ref{bern6}).

\subsection{Numerical Grid}
\label{num_grid}

The only symmetry assumption that can be made to simplify the problem
is equatorial symmetry across the $z = 0$ plane.  In the Newtonian
case, or in the case of maximal slicing $K=0$, the solutions are also
symmetric across the $y=0$ plane.  But the presence of a non-zero $K$
as in the case considered here there is no such symmetry condition
across the $y=0$ plane (see also Appendix E in \cite{ycsb04}).  The
numerical grid can therefore be chosen to cover a region $\hat x_{\rm
min} \leq \hat x \leq \hat x_{\rm max}$, $\hat y_{\rm min} \leq \hat y
\leq \hat y_{\rm max}$ and $0 \leq \hat z \leq \hat z_{\rm max}$.  For
the results in this paper we chose the limits to be the same in all
dimensions, $- \hat x_{\rm min} = \hat x_{\rm max} = - \hat y_{\rm
min} = \hat y_{\rm max} = \hat z_{\rm max} = \hat X_{\rm out}$.  In
this paper we also assume that $M_{\rm BH} \gg M_{\rm NS}$ and
restrict the numerical grid to a neighborhood of the neutron star,
which requires $\hat X_{\rm out} < \hat x_{\rm BH}$ (the origin of the
coordinate system is at the center of the neutron star, compare
Fig.~\ref{setup}).

\subsection{Elliptic Solver}

A variety of different approaches can be chosen to solve the elliptic
equations (\ref{Ham1}) -- (\ref{lapse1}), together with the algebraic
equation (\ref{bern6}).  We adopt finite difference methods, and solve
the resulting finite difference equations on uniform grids in a
parallel environment using DAGH software \cite{DAGH}.  Having adopted
a Kerr-Schild background, the operators in equations (\ref{Ham1}) --
(\ref{lapse1}) are not flat, which poses an additional challenge.  One
possible method of solving such a problem is to replace the
``covariant Laplace'' operator with a flat Laplace operator and solve
the equation iteratively.  In this paper we instead use PETSc software
\cite{PETSC} to solve linear elliptic equations of the form
\begin{equation} \label{ellsolver}
\hat D^2 f + u_f f = r_f
\end{equation}
directly.  Here $\hat D^2$ is a covariant Laplace operator
\begin{equation}
\hat D^2 f = \hat \gamma^{ij} f_{,ij} - \hat \Gamma^i f_{,i}
\end{equation}
and $u_f$ and $r_f$ are functions.  In the following we assume that
the metric $\hat \gamma^{ij}$ and the connection coefficients $\hat
\Gamma^i$ are given by the background quantities (\ref{KSsm}) and
(\ref{KScf}).  Equations (\ref{Ham1}) -- (\ref{lapse1}) can then be
solved by casting them into the form (\ref{ellsolver}).

\subsection{Hamiltonian Constraint}
\label{compham}

We write the conformal factor as a sum of contributions from
the black hole and the neutron star
\begin{equation} \label{split_psi}
\psi = \psi_{\rm BH} + \psi_{\rm NS}.
\end{equation}
In the absence of the neutron star the background metric 
$\hat \gamma_{ij}$ is already a solution to the constraints, so
\begin{equation}
\psi_{\rm BH} = 1
\end{equation}
identically.  The split (\ref{split_psi}) is convenient since
$\psi_{\rm NS}$ now satisfies a Robin fall-off conditions at large
separation from the neutron star.

Since the Hamiltonian constraint (\ref{Ham1}) contains non-linear
terms, the equation has to be linearized before it can be solved
with the linear operator (\ref{ellsolver}).  A solution to the
non-linear problem is then constructed iteratively.  We therefore
write the solution $\psi_{\rm NS}$ as a sum of the previous 
result $\Psi_{\rm NS}$ and a correction $\delta \psi_{\rm NS}$,
\begin{equation}
\psi_{\rm NS} = \Psi_{\rm NS} + \delta \psi_{\rm NS}.
\end{equation}
Inserting this into (\ref{split_psi}) and (\ref{Ham1}) then yields, to
first order in $\delta \psi_{\rm NS}$,
\begin{eqnarray} \label{psi_lin}
& &\hat D^2 \delta \psi_{\rm NS} - \\
& & \delta \psi_{\rm NS} \left(\frac{1}{8} \hat R
+ \frac{5}{12} \Psi^4 \hat K^2 
+ \frac{7}{8} \Psi^{-8} \hat A^2
- 10 \pi \Psi^4 \bar r_e^2 \bar \rho \right) \nonumber \\
& & = - \hat D^2 \Psi 
+ \frac{1}{8} \Psi \hat R + 
\frac{1}{12} \Psi^5 \hat K^2 - \frac{1}{8} \Psi^{-7} \hat A^2
- 2 \pi \Psi^5 \bar r_e^2 \bar \rho \nonumber
\end{eqnarray}
where we have used the abbreviations
\begin{equation}
\Psi = 1 + \Psi_{\rm NS}
\end{equation}
and
\begin{equation}
\hat A^2 = \hat A_{ij} \hat A^{ij}.
\end{equation}
Defining the residual of equation (\ref{Ham1}) as
\begin{equation}
\Res_{\psi} \equiv \hat D^2 \psi 
- \frac{1}{8} \psi \hat R - \frac{1}{12} \psi^5 \hat K^2 + 
\frac{1}{8} \psi^{-7} \hat A^2 + 2 \pi \psi^5 \bar r_e^2 
\bar \rho
\end{equation}
we can rewrite equation (\ref{psi_lin}) in the form (\ref{ellsolver})
with 
\begin{equation} \label{udeltapsi}
u_{\delta \psi_{\rm NS}} = - \left(\frac{1}{8} \hat R
+ \frac{5}{12} \Psi^4 \hat K^2 
+ \frac{7}{8} \Psi^{-8} \hat A^2
- 10 \pi \Psi^4 \bar r_e^2 \bar \rho \right)
\end{equation}
and
\begin{equation}
r_{\delta \psi_{\rm NS}} = - \Res_{\Psi}
\end{equation}
As discussed in Section \ref{reliter} we usually drop the linearized
matter term $10 \pi \Psi^4 \bar r_e^2 \bar \rho$ in equation
(\ref{udeltapsi}) above.  This does not affect the final solution, and
while removing this term slowed down the solution of the individual
equation, it did improve the convergence of the overall iteration
scheme.

In the above equations the background scalar curvature $\hat R$ is
given by
\begin{equation}
\hat R = 8 \frac{\alpha^4_{\rm BH} H^2}{\hat r_{\rm BH}^2},
\end{equation}
and $\hat A_{ij}$ is computed from equation (\ref{extcurv}) (see
Section \ref{compextcurv} below).

\subsection{The lapse equation}

It is convenient to introduce a new variable
\begin{equation}
N = \alpha \psi,
\end{equation}
in terms of which equation (\ref{lapse1}) can be rewritten as
\begin{eqnarray}\label{N}
\hat D^2 N & - &
\frac{7}{8} N \psi^{-8} \hat A^2
- \frac{5}{12} N \psi^4 \hat K^2
- \frac{1}{8} N \hat R
- \psi^5 \beta^i \hat D_i \hat K \nonumber \\
& = & 2 \pi N \psi^4 \bar r_e^2 (\bar \rho + 2 \bar S).
\end{eqnarray}
Similar to the split (\ref{split_psi}) we write $N$ as contributions
from the black hole and the neutron star
\begin{equation} \label{split_N}
N = N_{\rm BH} + N_{\rm NS}.
\end{equation}
In the absence of the neutron star $N_{\rm BH}$ reduces to
$\alpha_{\rm BH}$, so we may identify
\begin{equation}
N_{\rm BH} = \alpha_{\rm BH}
\end{equation}
The split (\ref{split_N}) can now be inserted into (\ref{N}), which
results in an equation of the form (\ref{ellsolver}) with
\begin{equation}
u_{N_{\rm NS}} = - \frac{7}{8} \psi^{-8} \hat A^2
- \frac{5}{12} \psi^4 \hat K^2
- \frac{1}{8} \hat R - 2 \pi \psi^4 \bar r_e^2 (\bar \rho + 2 \bar S)
\end{equation}
and
\begin{eqnarray}
r_{N_{\rm NS}} & = & - \hat D^2 \alpha_{\rm BH} 
+ \frac{7}{8} \alpha_{\rm BH} \psi^{-8} \hat A^2
+ \frac{5}{12} \alpha_{\rm BH} \psi^4 \hat K^2 \\
& & +  \frac{1}{8} \alpha_{\rm BH} \hat R + \psi^5 \beta^i \hat D_i K
+ 2 \pi \alpha_{\rm BH} \psi^4  \bar r_e^2 (\bar \rho + 2 \bar S) \nonumber
\end{eqnarray}
where
\begin{equation}
\hat D^2 \alpha_{\rm BH} = \frac{4 \alpha_{\rm BH}^7 H^2}{\hat r_{\rm BH}^2}.
\end{equation}

\subsection{The Momentum Constraint}

We first split the shift vector into contributions from the black hole,
the neutron star, and a rotational term
\begin{equation} \label{split_shift}
\beta^i = \beta^i_{\rm BH} + \beta^i_{\rm NS} + \xi^i,
\end{equation}
where
\begin{equation} \label{xi}
\xi^i = \epsilon^{ijk} \hat \Omega_j \hat x_k = 
\hat \Omega (-\hat y, \hat x - \hat x_{\rm rot}, 0).
\end{equation}
In equation (\ref{Mom1}), the analytic contributions can be moved
to the right hand side, leaving an equation for the neutron star
contribution
\begin{eqnarray} \label{Mom2}
\hat \Delta_L \beta^i_{\rm NS} & = &
2 \hat A^{ij} \hat D_j (\alpha \psi^{-6})
+ \frac{4}{3} \alpha \hat D^i \hat K \nonumber \\ 
& & + 16 \pi \alpha \psi^4 \bar r_e^2 \bar \jmath^i - 
\hat \Delta_L \beta^i_{\rm BH} - \hat \Delta_L \xi^i \nonumber \\
	& \equiv & S^i.
\end{eqnarray}
This equation can be solved with appropriate boundary conditions for
$\beta^i_{\rm NS}$ alone.  

The individual terms in (\ref{Mom2}) are computed as follows.  The 
first term on the right hand side is written as
\begin{eqnarray}
& & 2 \hat A^{ij} \hat D_j (\alpha \psi^{-6}) 
=  2 \hat A^{ij} \hat D_j (N \psi^{-7}) \\
& & =  2 \hat A^{ij} \psi^{-7} 
(\hat D_j \alpha_{\rm BH} + \hat D_j N_{\rm NS} - 7 N \psi^{-1} \hat D_j \psi).
\nonumber
\end{eqnarray}
Here the first term can be computed analytically from the Kerr-Schild
background
\begin{equation}
\hat D_j \alpha_{\rm BH} = \frac{\alpha_{\rm BH}^3 H}{\hat r_{\rm BH}} l_j.
\end{equation}
The gradient of $K$ in (\ref{Mom2}) is written as
\begin{eqnarray}
\frac{4}{3} \alpha \hat D^i \hat K  & = & \frac{4 N}{3 \psi} \hat D^i \hat K \\
& = & - \frac{\alpha_{\rm BH} + N_{\rm NS}}{\psi}\,
	\frac{8 \alpha_{\rm BH}^7 H}{3 \hat r_{\rm BH}^2}
	(2 + 10 H + 9 H^2) l^i. \nonumber
\end{eqnarray}
The vector Laplacian of $\beta^i_{\rm BH}$ is given by
\begin{equation}
\hat \Delta_L \beta^i_{\rm BH} = 
- \frac{8 \alpha_{\rm BH}^8 H}{3 \hat r_{\rm BH}^2} (2 + 12 H + 12 H^2) l^i.
\end{equation}
We note that in the absence of a neutron star, $N_{\rm NS} = \psi_{\rm
NS} = \hat \Omega = \bar \jmath^i = 0$, equation (\ref{Mom2}) is satisfied
analytically with $\beta^i_{\rm NS} = 0$.

The next goal is to bring the operator 
\begin{equation}
\hat \Delta_L \beta^i_{\rm NS} =
\hat D^2 \beta^i_{\rm NS} + \frac{1}{3} \hat D^i \hat D_j \beta_{\rm NS}^j
+ \hat R^i_{~j} \beta_{\rm NS}^j
\end{equation}
into the form (\ref{ellsolver}).  The mixed second derivatives 
can be eliminated by writing $\beta^i_{\rm NS}$ as the sum of 
a vector and the gradient of a scalar
\begin{equation}
\beta^i_{\rm NS} = W^i + \frac{1}{4} \hat D^i U.
\end{equation}
With this choice, equation (\ref{Mom2}) can be written as
\begin{equation} \label{U}
\hat D^2 U = - \hat D_i W^i
\end{equation}
and
\begin{equation} \label{W}
\hat D^2 W^i = S^i - \hat R^i_{~j} \left( W^j + \frac{1}{2} \hat D^j U \right).
\end{equation}
Since $U$ is a scalar, equation (\ref{U}) is already in the form
(\ref{ellsolver}), and can be solved immediately with
\begin{eqnarray}
u_U & = & 0 \\
r_U & = & - \hat D_i W^i.
\end{eqnarray}
Equation (\ref{W}) is more complicated.  Writing out the covariant
Laplace operator acting on a vector, we have
\begin{eqnarray} \label{covlap}
\hat D^2 W^l & = & \hat \gamma^{ij} W^l_{~,ij} - W^l_{~,k} \hat \Gamma^k 
	+ 2 \hat \gamma^{ij} W^m_{~,i} \hat \Gamma^l_{mj} \\
& & + W^m (\hat \gamma^{ij} \hat \Gamma^l_{mj,i}
	- \hat \Gamma^l_{mk} \hat \Gamma^k 
	+ \hat \gamma^{ij} \hat \Gamma^k_{mj} \hat \Gamma^l_{ki}). \nonumber
\end{eqnarray}
It is convenient to define
\begin{eqnarray}
& & \hat \Delta^l_{~m} \equiv \hat \gamma^{ij} \hat \Gamma^l_{mj,i}
	- \hat \Gamma^l_{mk} \hat \Gamma^k 
	+ \hat \gamma^{ij} \hat \Gamma^k_{mj} \hat \Gamma^l_{ki} \\
& & = \frac{2 \alpha_{\rm BH}^6 H}{\hat r_{\rm BH}^2}
\left( (1 + 2H)^2 \eta^l_{~m} - (3 + 13H+12H^2)l^l l_m \right) \nonumber
\end{eqnarray}
so that equation (\ref{covlap}) becomes
\begin{equation}
\hat D^2 W^l = \hat \gamma^{ij} W^l_{~,ij} - W^l_{~,k} \hat \Gamma^k
	+ 2 \hat \gamma^{ij} W^m_{~,i} \hat \Gamma^l_{mj} 
	+ W^m \hat \Delta^l_{~m}.
\end{equation}
To obtain the equation for $W^l$, the third term and the pieces of the
fourth term not containing $W^l$ can now be moved to the right hand
side of equation (\ref{W}).  The term $R^i_{~i} W^i$ (no summation) on
the right hand side of (\ref{W}) can also be absorbed in a $u_{W^i}$
on the left hand side, which leaves us with an equation for $W^i$ in
the form (\ref{ellsolver}) with
\begin{eqnarray}
u_{W^i} & = & \hat R^i_{~i} + \Delta^i_{~i} ~~~~~\mbox{no summation} \\
r_{W^i} & = & S^i - \hat R^i_{~j} W^j 
	-  \frac{1}{2} \hat R^i_{~l} \hat D^l U
	- 2 W^m_{~,l} \hat \Gamma^{il}_{m} \nonumber \\
& &	- W^j \Delta^i_{~j} ~~~~~ (j \neq i).
\end{eqnarray}
Here we have defined
\begin{eqnarray}
\hat \Gamma^{li}_{m} & \equiv & \hat \gamma^{ij} \hat \Gamma^l_{mj} \\
& = & \frac{\alpha_{\rm BH}^4 H}{\hat r_{\rm BH}}
\left(2 (1 + 2 H) l^l \eta^i_{~m} - (3 + 4H)l^l l^i l_m \right). \nonumber
\end{eqnarray}
We also note that 
\begin{eqnarray}
& & \hat R^i_{~j} = \\ 
& & \frac{\alpha^6_{\rm BH} H}{\hat r^2_{\rm BH}}
\left( (1 + 6H + 8 H^2) \eta^i_{~j} - (3 + 10H + 8H^2) l^i l_j \right).
\nonumber
\end{eqnarray}

In $S^i$, $\hat \Delta_L \xi^i$ can be computed from (\ref{veclap})
and (\ref{covlap}).  The partial derivatives can be computed trivially
from (\ref{xi}); in particular we have $\xi^l_{,ij} = 0$.  Because of
the symmetries we also have $\xi^m_{~,i} \hat \Gamma^{li}_{m} = 0$,
which simplifies the calculation.  For extreme mass ratios, when
$x_{\rm rot} = x_{\rm BH}$, $\hat \Delta_L \xi^i = 0$ identically.
Equation (\ref{W}) can now be solved iteratively for the components
$W^i$.

\subsection{The Extrinsic Curvature}
\label{compextcurv}

With a new solution for the shift $\beta^i$, the extrinsic curvature
$\hat A^{ij}$ can be computed from equation (\ref{extcurv}).  In
this expression, the shift decomposition (\ref{split_shift}) has
to be inserted.  The background contribution of $\beta^i_{\rm BH}$
can be computed analytically
\begin{eqnarray}
& & (\hat L \beta_{\rm BH})^{ij} = \\
& & \frac{4 \alpha^6_{\rm BH} H}{3 \hat r_{\rm BH}}
    \left( (2 + 7H + 6H^2) \eta^{ij} - ( 6 + 13H + 6 H^2) l^i l^j \right).
\nonumber
\end{eqnarray}
The contributions of the rotational term $\xi^i$ are most easily added
by calculating the partial derivatives directly from (\ref{xi}), and
computing the connection terms together with those for the neutron
star terms $\beta^i_{\rm NS}$.

\subsection{The integrated Euler equation}

As in the Newtonian case we first evaluate the integrated Euler
equation (\ref{bern5}) at the three points $\hat x_A$, $\hat x_B$ and
$\hat x_C$ to find the three eigenvalues $C$, $\hat \Omega$ and $\bar
r_e$.  All quantities in the integrated Euler equation have to be
expressed in terms of our decompositions (\ref{split_psi}),
(\ref{split_N}) and (\ref{split_shift}).  Of the three constants, only
$C$ appears directly in (\ref{bern5}).  The angular velocity $\hat
\Omega$ has also been introduced explicitly through the rotational
shift term $\xi^i$ in (\ref{split_shift}).  The scaling parameter
$\bar r_e$ again only appears through the implicit scaling of the
gravitational potentials.  In the Newtonian case, the Poisson equation
(\ref{poisson3}) immediately implied a scaling for $\phi_{\rm NS}$.
The corresponding relativistic field equations, namely the Hamiltonian
constraint (\ref{Ham1}) and the lapse equation (\ref{lapse1}), do not
provide such a simple scaling because the non-linear terms.  However,
we can use the Newtonian limit to guess an appropriate scaling
(compare \cite{bcsst}.)

In the Newtonian limit, the lapse becomes
\begin{equation}
\alpha \sim e^{\phi} \sim 1 + \phi = 1 + \phi_{\rm NS} + \phi_{\rm BH}.
\end{equation}
Since the background describes a black hole, the conformal factor
only includes a neutron star contributions, so that in the Newtonian
limit
\begin{equation}
\psi \sim e^{-\phi_{\rm NS}/2} \sim 1 - \frac{1}{2} \phi_{\rm NS}.
\end{equation}
Comparing with (\ref{split_psi}) we note that in the Newtonian 
limit we have
\begin{equation}
\psi_{\rm NS} \sim  - \frac{1}{2} \phi_{\rm NS}
\end{equation}
and from (\ref{split_N})
\begin{equation}
N_{\rm NS} = \alpha \psi - \alpha_{\rm BH} \sim \frac{1}{2} \phi_{\rm NS}
\end{equation}
In analogy with the Newtonian problem, this suggests the definition
\begin{equation}
\hat \psi_{\rm NS} = \psi_{\rm NS}/\bar r_e^2 ~~~~~~~~~
\hat N_{\rm NS} = N_{\rm NS}/\bar r_e^2
\end{equation}
In the integrated Euler equation, $\psi_{\rm NS}$ and $N_{\rm NS}$ can 
then be replaced with $\bar r_e^2 \hat \psi_{\rm NS}$ and 
$\bar r_e^2 \hat N_{\rm NS}$, which explicitely introduces the 
scaling parameter $\bar r_e$ into the equation.  

Once the three eigenvalues  $C$, $\hat \Omega$ and $\bar r_e$ have
been found,  $\psi_{\rm NS}$ and $N_{\rm NS}$ have to be rescaled to
reflect the new value of $\bar r_e$.  The integrated Euler equation can
then be solved everywhere for the new density distribution $q$.

\subsection{Boundary Conditions}
\label{BC}

In general, the construction of BHNS binaries requires three different
kinds of boundary conditions: exterior boundary conditions at large
separation from the binary, inner boundary conditions on the black
hole's excision surface, and, having adopted equatorial symmetry (see
Section \ref{num_grid}), a symmetry boundary condition on the $z=0$
plane.  The symmetry conditions on the $z=0$ are straight-forward to
implement (see Table \ref{BC_table}).  As explained in Section
\ref{sec4}, the assumption of extreme mass ratios $M_{\rm BH} \gg
M_{\rm NS}$ allows us to restrict the numerical grid to a region
around the neutron star.  This eliminates the need for interior
excision boundary conditions, but it also requires imposing boundary
conditions in the potentially strong-field regime close to the black
hole.

Given that we expect the neutron star to affect the spacetime only in
a small neighborhood, it is reasonable to assume $1/r$ Robin fall-off
conditions for the conformal factor $\psi_{\rm NS}$ and lapse $N_{\rm
NS}$.  

\begin{table}
\begin{center}
\begin{tabular}{c|cc}
& $ z = 0$ & OB \\
\tableline 
$\psi_{\rm NS}$ & SYM  & $\sim \displaystyle \frac{1}{r}$ \\[3mm] 
$N_{\rm NS}$ 	& SYM  & $\sim \displaystyle \frac{1}{r}$ \\[3mm]
$U$ 		& SYM  & $\sim \displaystyle \frac{y}{r}$ \\[3mm]
$W^x$ 		& SYM  & $\sim \displaystyle \frac{1}{r}$ \\[3mm]
$W^y$ 		& SYM  & $\sim \displaystyle \frac{1}{r}$ \\[3mm]
$W^z$ 		& ~~ANTI~~  & $\sim \displaystyle \frac{z}{r^3}$ \\[3mm]

\end{tabular}
\end{center}
\caption{Boundary conditions for the outer boundaries (OB) and on the
$z=0$ symmetry plane (where SYM denotes symmetric and ANTI
antisymmetric boundary conditions).  Note the $r$ is the separation
from the origin of the coordinate system at the center of the neutron
star.}
\label{BC_table}
\end{table}

\begin{table}
\begin{center}
\begin{tabular}{cccccc}
\tableline \tableline
$\hat X_{\rm out}$ & Grid size & $x_{\rm BH}/M_{\rm BH}$ & $r_e$ & 
$q_{\rm max}$ & $\Omega M_{\rm BH}$ \\
\tableline 
2 & $(32 \times 32 \times 16)$ & 17.82  &   1.113 &   0.02349 &  0.01330 \\
3 & $(48 \times 48 \times 24)$ & 17.81	&   1.113 &   0.02349 &  0.01332 \\
4 & $(64 \times 64 \times 32)$ & 17.81  &   1.113 &   0.02349 &  0.01333 \\
6 & $(96 \times 96 \times 48)$ & 17.81  &   1.113 &   0.02349 &  0.01336 \\
\tableline \tableline
\end{tabular}
\end{center}
\caption{Numerical results for different locations of the outer
boundary at constant grid resolution.  We show values for a $\bar
M_{\rm BH} = 0.5$, $\bar M_{\rm NS} = 0.05$ binary at a separation of
$\hat x_{\rm BH} = -8$ (see also the top line in Table
\ref{Seq_table}).}
\label{Conv_table}
\end{table}

More subtle are the boundary conditions on the shift quantities $W^i$
and $U$.  In \cite{bcsst} a similar decomposition was used for the
construction of binary neutron stars, and boundary conditions for
these fields were identified from the asymptotic behavior of a
multipole solution to a general, but flat vector Laplacian \cite{b82}.
The construction also assumed that the net momentum contained in the
numerical grid vanishes (so that the non-vanishing angular momentum
gives rise to the asymptotic fall-off.)  Here, however, we solve a
vector Laplacian that is not flat, we adopt boundary conditions which
are not in an asymptotic region, and the grid containing only the
neutron star contains a non-vanishing linear momentum.  It is not
clear a priori how appropriate boundary conditions on $W^i$ and $U$
should be constructed in this case.  We have experimented with several
different conditions, and did notice some effect on the convergence
properties of the iteration scheme.  We found reasonable results with
the conditions tabulated in Table \ref{BC_table}, which are motivated
by the construction in \cite{b82} but do not assume a vanishing linear
momentum.  In Table \ref{Conv_table} we tabulate numerical values
obtained for different locations of the outer boundaries, which
demonstrates that our results depend only very weakly on where the
outer boundaries are imposed (as long as they are imposed sufficiently
far away away from the neutron star and not too close to the black
hole singularity).  Once the assumption of extreme mass ratios is
relaxed and the computational grid encompasses both the neutron star
and the black hole, the boundary conditions of \cite{bcsst} can again
be applied.

\subsection{Iteration Scheme}
\label{reliter}

The iteration scheme used for the construction of relativistic BHNS
binaries is very similar to that for Newtonian binaries described in
Section \ref{Niter}.  However, instead of having to solve only one
Poisson equation, we now have to solve several Poisson-like equations
together with the integrated Euler equation.  These different
equations can be solved in various different sequences.  We found best
convergence when solving the integrated Euler equation after each
elliptic equation.

We also found that the convergence improved by using underrelaxation,
i.e.~by using a linear combination 
\begin{equation}
f^{n+1} = \chi \tilde f^{n+1} + (1 - \chi) f^n,
\end{equation}
where $f^n$ is any one of the functions $N_{\rm NS}$, $\psi_{\rm NS}$,
$W^i$ and $U$ after $n$ iteration steps, and $\tilde f^{n+1}$ the
function after having solved the corresponding elliptic equation.  The
parameter $\chi$ is the relaxation parameter, and values of $\chi < 1$
define underrelaxation.  We have found best results for $\chi$ between
0.6 and 0.8.  

While the Hamiltonian constraint alone converged faster when solved as
described in Section \ref{compham}, we also found improved convergence
of the overall iteration when we did not include the linearized matter
term in equation (\ref{udeltapsi}).  The effect of not including this
term is similar to that of using underrelaxation, and reduced the
coupling between the matter and conformal factor, which otherwise led
to unstable behavior in some situations.

\end{appendix}


\begin{thebibliography}{99}
\bibitem{visc_fn} We point out that \cite{bc92} do not consider the
extreme mass ratio limit.  However, from their equation (8) it is
clear that extreme mass ratios lead to an even more severe constraint,
making it even more unlikely for neutron stars to be corotational.
Physically, tidal effects on the neutron star are less important for
much larger companion masses; similarly, the neutron star tidal
dissipation timescale increases even more rapidly than the inspiral
timescale as the black hole-to-neutron star mass ratio increases.
\bibitem{fn1} In the neighborhood of the black hole, deviations of the
conformal factor $\psi$ from unity are in the order of $M_{\rm
NS}/x_{\rm BH}$.  Assuming that the fractional deviation of irreducible
mass from the background mass $M_{\rm BH}$ is proportional to the
deviation of the conformal factor from unity, $(M_{\rm irr} - M_{\rm
BH})/M_{\rm BH} \propto \psi - 1 \propto M_{\rm NS}/x_{\rm BH}$, then
implies that the absolute deviation is proportional to the 
binding energy $M_{\rm BH} M_{\rm NS}/x_{\rm BH}$.
\end{thebibliography}
\end{document}